\documentclass{iopart}

\usepackage{graphicx}  
\usepackage{latexsym}
\usepackage{amssymb}

\jl{6}        
\eqnobysec    

\def\beq{\begin{equation}}
\def\eeq{\end{equation}}
\def\rmd{{\rm d}} 

\begin{document}

\title[On the energy content of electromagnetic and gravitational plane waves]
{On the energy content of electromagnetic and gravitational plane waves through super-energy tensors}

\author{
Donato Bini${}^*{}^\S{}^\P$
and 
Andrea Geralico${}^*$
}

\address{
  ${}^*$\
  Istituto per le Applicazioni del Calcolo ``M. Picone,'' CNR, Via dei Taurini 19, I-00185, Rome, Italy
}
\address{
  ${}^\S$\
  ICRANet, Piazza della Repubblica 10, I-65122 Pescara, Italy
}
\address{
  ${}^\P$
  INFN, Sezione di Napoli, Via Cintia, Edificio 6 - 80126 Napoli, Italy 
}

\ead{donato.bini@gmail.com}
\ead{andrea.geralico@gmail.com}

\begin{abstract}
The energy content of (exact) electromagnetic and  gravitational plane waves is studied in terms of super-energy tensors (the Bel, Bel-Robinson and the --less familiar-- Chevreton tensors) and natural observers.
Starting from the case of single waves, the more interesting situation of colliding waves is then discussed, where the nonlinearities of the Einstein's theory play an important role.
The causality properties of the super-momentum four vectors associated with each of these tensors are also investigated when passing from the single-wave regions to the interaction region.
\end{abstract}

\pacno{04.20.Cv}

\section{Introduction}

Electromagnetic and gravitational waves are the most important features associated with electromagnetic and gravitational phenomena.
While in a flat (special relativistic) spacetime context the energy content of an electromagnetic wave is well defined, in the curved (general relativistic) spacetime situation  there exist different definitions bearing either a more physical meaning or a more direct geometrical one, and a debated question is on what should be preferred or discarded.

At the level of the full nonlinear theory, strong electromagnetic waves are exact solutions of the Einstein-Maxwell field equations obeying the standard symmetries of a classical electromagnetic wave in flat spacetime. Similarly, strong gravitational waves are exact solutions of the vacuum Einstein's equations sharing (only) the symmetries of the electromagnetic waves. 
Their collision in a curved spacetime represents a completely different situation with respect to the flat spacetime. In fact, in this case either the electromagnetic waves or the gravitational waves \lq\lq gravitate" themselves, and the result of their mutual interaction is not a simple linear superposition of the associated fields, but rather a nonlinear superposition which may end with a focusing of the incoming waves, and the subsequent creation of a spacetime singularity
(or sometimes with the creation of a Killing-Cauchy horizon).
Working with exact solutions, simplifications arise  when considering the case of plane electromagnetic and gravitational waves undergoing head-on collision.
In this situation (to which we will limit our considerations) the spacetime representing a collision problem is actually a puzzle of matching pieces, namely a flat spacetime region (before the passage of the waves), two Petrov type N regions (the two incoming single waves, undergoing then a head-on collision) and a type I (or less general, e.g., type D) spacetime region corresponding to the interaction zone (see, e.g., Refs. \cite{Griffiths:1991zp,Chandrasekhar:1991} for a detailed account).

The spacetime curvature associated with either electromagnetic or gravitational waves induces observable effects on the motion of test particles and photons.
For example, one can study the scattering of massive and massless neutral scalar particles (in motion along geodesics) by plane gravitational waves (see Ref. \cite{Garriga:1990dp}, where both the classical and quantum regimes of the scattering were considered).
One can also consider the nongeodesic motion of massive particles interacting with an electromagnetic pulse and accelerated by the radiation field.
In fact, during the scattering process the particle may absorb and re-emit radiation as a secondary effect, resulting in a force term acting on the particle itself \cite{Bini:2014uoa}. 
This said, it is a matter of fact that test particles might interact very differently with either a gravitational wave or an electromagnetic wave background. The observables associated with the scattering process, like the cross section, will then contain a signature of the different nature of the host environment \cite{Bini:2012gc}.
Similarly, the scattering of light by the radiation field given by either an exact gravitational plane wave or an exact electromagnetic wave or their collision will present specific and different features \cite{Bini:2014ihf,Bini:2014rga}.

In this work we aim at characterizing single and colliding electromagnetic and gravitational plane wave spacetimes in terms of super-energy tensors with respect to natural observers and frames. 
The notion of a super-energy tensor was suggested long ago by Bel \cite{bel58,bel59,bel62} and Robinson \cite{rob59,rob97} for general gravitational fields, in order to provide a definition of (local) energy density and energy flux in complete formal analogy with electromagnetism.
The properties of the Bel and Bel-Robinson tensors have been extensively investigated over the years \cite{Deser:1977zz,Mashhoon:1996wa,Bonilla:1997,Bergqvist:1998a,Bergqvist:1998b,Mashhoon:1998tt,Senovilla:1999qv,Bonilla:1999in,Senovilla:1999xz,Senovilla:1999xs,bini-jan-miniutti,Lazkoz:2003wx,Bergqvist:2004qh,GomezLobo:2007gm,Ferrando:2008sr,Gomez-Lobo:2013eya,Clifton:2014lha}, 
so that they are now considered as a useful mathematical tool to describe the energy content of a given spacetime. 
In particular, using the canonical expressions of the Weyl tensor associated with different Petrov type spacetimes, Bel himself showed \cite{bel62}
that for Petrov types I and D, an observer always exists for 
which the spatial super-momentum density vanishes. In addition, this observer is peculiar enough because he aligns the 
electric and magnetic parts of the Weyl tensor in the sense that they are
both diagonalized (and commuting). 
For black hole spacetimes, the Carter observer family plays exactly this role, but not much literature exists in spacetimes different from black holes \cite{bini-jan-miniutti}. 
Despite their local character, the Bel and Bel-Robinson tensors have also been used to investigate the global stability properties of Minkowski spacetime \cite{Christodoulou:1993uv}. 

A third super-energy tensor has received attention in the last years, the Chevreton tensor.
It was introduced by Chevreton \cite{Chevreton:1964} for the Maxwell field in close analogy with the Bel-Robinson tensor, in the sense that it has been sought for depending on the second derivatives of the electromagnetic potential (first derivatives of the Faraday tensor), like the Bel-Robinson tensor which involves the Weyl (or Riemann) tensor, and hence the second derivatives of the metric (i.e., the gravitational potential).
General properties of the Chevreton tensor and conservation laws have been studied only recently in Refs. \cite{Bergqvist:2003an,Senovilla:2003ba,Edgar:2003gc,Eriksson:2005ab,Bergqvist:2007hi,Eriksson:2007vx,Lazkoz:2003wx} with some applications to explicit spacetime solutions. 

We will investigate the energy content of some simple spacetimes associated with exact electromagnetic and gravitational plane waves and their collision in terms of the super-energy density and the super-Poynting vector built with the above super-energy tensors \cite{Maartens:1997fg,bini-jan-miniutti}.
We select natural observer-adapted frames to evaluate the corresponding super-momentum four vector, which carries the twofold information of the super-energy density and the linear super-momentum density.
The latter is an observer-dependent quantity by definition, but has the advantage to summarize (and control during their evolution) both super-energy and super-momentum densities region-by-region.
 
The paper  is organized as follows. In Sec. \ref{BR} we will shortly revisit the main properties of the Bel and Bel-Robinson tensors, and those 
of the Chevreton tensor. We introduce then the metrics of a single electromagnetic and gravitational wave in Sec. 3 and compare among them for what concerns the energy (and momentum) content.
Finally, in Sec. 4, we extend our considerations to the problem of collision, discussing then in Sec. 5 how the nonlinearities of the theory act on the energetics. 
  
We will use geometrical units and conventionally assume that Greek indices run from $0$ to $3$, whereas Latin indices run from $1$ to $3$. The metric signature is chosen to be $-+++$. When necessary  spacetime splitting techniques will be used to express tensors in the associated $1+3$ form, following    Refs. \cite{Jantzen:1992rg,Felice:2010cra} for notations and conventions.

\section{Super-energy-momentum tensors }
\label{BR}

We recall below the definition of the Bel and Bel-Robinson tensors and of the Chevreton tensor.
These tensors are referred to as super-energy tensors, since they share similar properties to the ordinary energy-momentum tensors, even if they do not have units of an energy per unit of spatial volume, but rather of an energy per unit surface.
The Bel and Bel-Robinson tensors are built with the Riemann tensor and the Weyl tensor, respectively, and are both divergence-free in vacuum spacetimes, where they coincide.
The Chevreton tensor is an electromagnetic counterpart of the Bel-Robinson tensor. It is quadratic in the covariant derivatives of the Faraday tensor, and is not divergence-free even in absence of electromagnetic sources.
It is however divergence-free in flat spacetime \cite{Senovilla:1999xz}.
These tensors all satisfy a positive-definite property, i.e., their full contraction with any four future-pointing vectors is always non-negative. 
This was known for the Bel-Robinson tensor, but only proved much later for the Bel, and the Chevreton tensors \cite{Senovilla:1999xz}.

\subsection{Bel and Bel-Robinson tensors}

The Bel and Bel-Robinson tensors \cite{bel58,bel59,bel62,rob59,rob97} are defined as 
\begin{eqnarray}
\label{Bel}
 T_{\rm B}^{\rm (g)}{}_{\alpha\beta} {}^{\gamma\delta}
 &=& k
   ( R_{\alpha}{}^{\rho\gamma\sigma} R_{\beta\rho}{}^{\delta}{}_{\sigma}
   + {}^* R_{\alpha}{}^{\rho\gamma\sigma} {}^* R_{\beta\rho}{}^{\delta}{}_{\sigma}\nonumber\\
&& 
+ R^*{}_{\alpha}{}^{\rho\gamma\sigma} R^*{}_{\beta\rho}{}^{\delta}{}_{\sigma}
+ {}^* R^*{}_{\alpha}{}^{\rho\gamma\sigma} {}^* R^*{}_{\beta\rho}{}^{\delta}{}_{\sigma})
\ ,
\end{eqnarray}
in terms of the Riemann tensor  $R_{\alpha\beta\gamma\delta}$, and 
\beq
\label{BelR}
 T_{\rm BR}^{\rm (g)}{}_{\alpha\beta} {}^{\gamma\delta}
 = 2k
   ( C_{\alpha\rho\beta\sigma} C^{\gamma\rho\delta\sigma}
   + {}^* C_{\alpha\rho\beta\sigma} {}^* C^{\gamma\rho\delta\sigma} )
\ ,
\eeq
in terms of the Weyl tensor $C_{\alpha\beta\gamma\delta}$, respectively.
Their mathematical properties with associated proofs are reviewed, e.g., in Refs. \cite{Bonilla:1997,Senovilla:1999xz}.
They are a natural generalization\footnote{
The prefactor $k$ can have different values, according to different definitions associated or not 
with the number of contracted indices. We will assume $k=1/2$ hereafter, following the original definition by Bel \cite{bel58}.
} 
of the electromagnetic energy-momentum tensor, which reads
\beq
 T^{\rm (em)}{}_\alpha{}^\beta
 = \frac{1}{2}
   ( F_{\alpha\rho} F^{\beta\rho}
   + {}^* F_{\alpha\rho} {}^* F^{\beta\rho} )\,.
\eeq

Similarly, in direct analogy with
the (observer-dependent) definition of electromagnetic energy density and Poynting
vector with respect to a given observer $u$ (with $u\cdot u=-1$, $u$ denoting the future-pointing unit tangent vector to his/her world lines and $P(u)=g+u\otimes u$ the projection map  orthogonally to $u$), i.e., 
\beq
  \mathcal{E}^{\rm (em)}(u) =T^{\rm (em)}_{\alpha\beta} u^\alpha u^\beta\ ,\qquad
  \mathcal{P}^{\rm (em)}(u)_\alpha = P(u)^\epsilon{}_\alpha T^{\rm (em)}_{\epsilon\beta} u^\beta\ ,
\eeq
so that the following four vector
\beq
{\sf P}^{\rm (em)}(u)=  -\mathcal{E}^{(\rm{em})}(u) u +\mathcal{P}^{(\rm{em})} (u)\equiv [T^{\rm (em)}{}_{\gamma\delta} u^\delta]\,,
\eeq
can be naturally formed, 
the super-energy density and the super-momentum density (or super-Poynting, spatial) vector \cite{bel58,bel59,bel62,rob59} (see also Refs. \cite{Maartens:1997fg,bini-jan-miniutti}) are obtained from the Bel and Bel-Robinson tensors by
\begin{eqnarray}
\label{Belsuperendef}
 \mathcal{E}^{\rm{(g)}}_{\rm B}(u)
 &=&  T^{\rm (g)}_{\rm B}{}_{\alpha\beta\gamma\delta} u^\alpha u^\beta u^\gamma u^\delta\ ,
\nonumber\\
 \mathcal{P}^{\rm{(g)}}_{\rm B} (u)_\alpha
 &=& P(u)^\epsilon{}_\alpha T^{\rm (g)}_{\rm B}{}_{\epsilon\beta\gamma\delta} u^\beta u^\gamma u^\delta
\,,
\end{eqnarray}
and
\begin{eqnarray}
\label{BelRsuperendef}
 \mathcal{E}^{\rm{(g)}}_{\rm BR}(u)
 &=&  T^{\rm (g)}_{\rm BR}{}_{\alpha\beta\gamma\delta} u^\alpha u^\beta u^\gamma u^\delta\ ,
\nonumber\\
 &=&  2k {\rm Tr}\, [ E(u)\cdot E(u) + H(u)\cdot H(u) ]\ ,\nonumber\\
 \mathcal{P}^{\rm{(g)}}_{\rm BR} (u)_\alpha
 &=& P(u)^\epsilon{}_\alpha T^{\rm (g)}_{\rm BR}{}_{\epsilon\beta\gamma\delta} u^\beta u^\gamma u^\delta
\nonumber\\
 &=&  4k[ E(u)\times_u H(u)]_\alpha\,,
\end{eqnarray}
respectively.
We denote by ${\mathcal E}(u)$, ${\mathcal H}(u)$ and ${\mathcal F}(u)$ the electric, magnetic and mixed parts of the Riemann tensor relative to the observer $u$ \cite{Jantzen:1992rg} 
\beq\fl\quad
{\mathcal E}(u)_{\alpha\beta}=R_{\alpha\mu\beta\nu}u^\mu u^\nu\,,\quad 
{\mathcal H}(u)_{\alpha\beta}={}^*R_{\alpha\mu\beta\nu}u^\mu u^\nu\,,\quad
{\mathcal F}(u)_{\alpha\beta}={}^*R^*{}_{\alpha\mu\beta\nu}u^\mu u^\nu\,,
\eeq
and by $E(u)$ and $H(u)$ the electric and magnetic parts of the Weyl tensor 
 \beq
\label{EHweyl}
 E(u)_{\alpha\beta}=C_{\alpha\mu\beta\nu}u^\mu u^\nu\,,\qquad 
H(u)_{\alpha\beta}={}^*C_{\alpha\mu\beta\nu}u^\mu u^\nu\,.
 \eeq
Furthermore, $[A\times_u B]_{\alpha} = \eta (u)_{\alpha\beta\gamma} A^{\beta} {}_{\delta}\, B^{\delta\gamma}$ defines a spatial cross product for two symmetric spatial tensor fields ($A, B$), with $\eta (u)_{\alpha\beta\gamma} 
= u^\delta \,\eta_{\delta\alpha\beta\gamma}$ ($\eta_{\delta\alpha\beta\gamma}=\sqrt{-g}\epsilon_{\delta\alpha\beta\gamma}$, $\epsilon_{0123}=1$) the unit volume 3-form.
The orthogonal splitting of the Bel and Bel-Robinson tensors is discussed in Appendix (see also Ref. \cite{GomezLobo:2007gm}).

The following four vector \cite{Senovilla:1999xz}
\beq
{\sf P}^{\rm (g)}(u)=  -\mathcal{E}^{(\rm{g})}(u) u +\mathcal{P}^{(\rm{g})}(u)\equiv [T^{\rm (g)}{}_{\epsilon\beta\gamma\delta} u^\beta u^\gamma u^\delta]\,,
\eeq
can then be naturally formed using either the Bel or Bel-Robinson tensor, as in the electromagnetic case. 
When ${\sf P}^{\rm (g)}(u)$ is not a null vector, it can also be rewritten as
\begin{eqnarray}
{\sf P}^{\rm (g)}(u)&=&  -\mathcal{E}^{(\rm{g})}(u) u +||\mathcal{P}^{(\rm{g})}(u)||\hat\mathcal{P}^{(\rm{g})}(u)\nonumber\\
&\equiv&-\sqrt{[\mathcal{E}^{(\rm{g})}(u)]^2-||\mathcal{P}^{(\rm{g})}(u)||^2}\, {\sf U}^{(\rm{g})}\,,
\end{eqnarray}
where $\hat\mathcal{P}^{(\rm{g})}(u)$ is a unit spatial vector and the following unit timelike vector
\beq
{\sf U}^{(\rm{g})}=\frac{\mathcal{E}^{(\rm{g})}(u)}{\sqrt{[\mathcal{E}^{(\rm{g})}(u)]^2-||\mathcal{P}^{(\rm{g})}(u)||^2}}\left[
u -{\mathcal V}({\sf U}^{\rm (g)},u)\hat\mathcal{P}^{(\rm{g})}(u)
\right]\,,
\eeq
aligned with the super-momentum four vector has been introduced.
The quantity 
\beq
\label{calVdef}
{\mathcal V}({\sf U}^{\rm (g)},u)=\frac{||\mathcal{P}^{(\rm{g})}(u)||}{\mathcal{E}^{(\rm{g})}(u)}\,,
\eeq
denotes the relative velocity of ${\sf U}^{\rm (g)}$ with respect to $u$.
When instead ${\sf P}^{\rm (g)}(u)$ is   a null vector, then
\beq
\frac{||\mathcal{P}^{(\rm{g})}(u)||}{\mathcal{E}^{(\rm{g})}(u)}=\pm 1\,,
\eeq
and one has the representation
\beq
{\sf P}^{\rm (g)}(u)=  -\mathcal{E}^{(\rm{g})}(u) [ u \pm \hat\mathcal{P}^{(\rm{g})}(u)]\,.
\eeq

Let us mention that the electromagnetic and gravitational four vectors ${\sf P}^{\rm (em)}(u)$ and ${\sf P}^{\rm (g)}(u)$ so defined are past directed, since the super-energy densities $\mathcal{E}^{(\rm{em})}(u)$ and $ \mathcal{E}^{\rm{(g)}}(u)$ are positive definite.
Furthermore, they are observer-dependent quantities and transform  consequently when passing from one observer to another with four velocities related by a boost, e.g.,
\beq
U=\gamma(U,u)[u+\nu(U,u)]\,,
\eeq
where $\gamma(U,u)$ is the Lorentz factor and $\nu(U,u)$ is the relative velocity of $U$ with respect to $u$, i.e., of the new observer with respect to the first one.

\subsection{Chevreton tensor}
\label{Chev}

The Chevreton tensor \cite{Chevreton:1964} is defined by 
\beq
\label{chev}
H_{\alpha\beta\lambda\mu}
 = \frac{1}{2}
   ( E_{\alpha\beta\lambda\mu}+ E_{\lambda\mu\alpha\beta})
\ ,
\eeq
with
\beq
E_{\alpha\beta\lambda\mu}=Y_{\alpha\lambda\beta\mu}+Y_{\alpha\mu\beta\lambda}-g_{\alpha\beta}X_{\lambda\mu}-\frac12g_{\lambda\mu}Z_{\alpha\beta}+\frac14g_{\alpha\beta}g_{\lambda\mu}Y\,,
\eeq
where we have used the notation
\beq\fl\qquad
Y_{\alpha\beta\lambda\mu}=\nabla_\alpha F_{\beta\rho} \nabla_\lambda F_\mu{}^\rho\,,\quad
X_{\lambda\mu}=Y_{\sigma\lambda}{}^\sigma{}_{\mu}\,,\quad
Z_{\alpha\beta}=Y_{\alpha\sigma\beta}{}^\sigma\,,\quad
Y=Y_{\tau\sigma}{}^{\tau\sigma}\,.
\eeq
It has the algebraic symmetries
\beq
H_{\alpha\beta\lambda\mu}=H_{(\alpha\beta)(\lambda\mu)}=H_{\lambda\mu\alpha\beta}=H_{(\alpha\beta\lambda\mu)}\,.
\eeq
The last property holds in four dimensions only and in the absence of electromagnetic sources \cite{Bergqvist:2003an}.

One can define a super-energy density and a super-Poynting vector also in this case
\begin{eqnarray}
 \mathcal{E}^{\rm{(g)}}_{\rm C}(u)
 &=&  H_{\alpha\beta\gamma\delta} u^\alpha u^\beta u^\gamma u^\delta\ ,
\nonumber\\
 \mathcal{P}^{\rm{(g)}}_{\rm C} (u)_\alpha
 &=& P(u)^\epsilon{}_\alpha H_{\epsilon\beta\gamma\delta} u^\beta u^\gamma u^\delta
\,,
\end{eqnarray}
leading to the Chevreton super-momentum four vector
\beq
{\sf P}^{\rm (g)}_{\rm C}(u)=  -\mathcal{E}^{(\rm{g})}_{\rm C}(u) u +\mathcal{P}^{(\rm{g})}_{\rm C}(u)\equiv [H_{\epsilon\beta\gamma\delta} u^\beta u^\gamma u^\delta]\,.
\eeq

These super-momentum four vectors, namely ${\sf P}^{\rm (g)}_{\rm C}(u)$ and ${\sf P}^{\rm (g)}(u)$ (in both cases of Bel and Bel-Robinson gravitational super-energy-momentum tensors) will be analyzed below, in a context of electromagnetic and gravitational wave propagation.

\section{Single plane waves}

The gravitational field associated with (exact) electromagnetic and gravitational plane waves with a single polarization
state is described by the line element (see, e.g., Refs. \cite{Bell:1974vb,Griffiths:1975zm})
\beq
\label{radfield}
\rmd s^2 = - 2\rmd u \rmd v +F^2(u)\rmd x^2+G^2(u)\rmd y^2\,, 
\eeq
written in  coordinates $x^\alpha=(u,v,x,y)$ adapted to the wave front.
The null coordinates $(u,v)$ can be related to standard Cartesian coordinates $(t,z)$ by the transformation
\beq
\label{u_coord}
u=\frac{1}{\sqrt{2}}(t-z)\,,\qquad v=\frac{1}{\sqrt{2}}(t+z)\,.
\eeq
Referred to the coordinates $(t,x,y,z)$, Eq. (\ref{radfield}) assumes a quasi-Cartesian form 
\beq
\label{quasicart}
\rmd s^2 = -\rmd t^2 + F^2(t-z)\rmd x^2+G^2(t-z)\rmd y^2+ \rmd z^2\,,
\eeq
and 
\beq
\partial_t =\frac{1}{\sqrt{2}}(\partial_u+\partial_v) \,,\quad \partial_z=\frac{1}{\sqrt{2}}(-\partial_u+\partial_v) \,,
\eeq
so that the waves are traveling along the $z$-axis, while $x$ and $y$ are two spacelike coordinates on the wave front. 
 
A family of fiducial observers at rest with respect to the coordinates $(x,y,z)$ is characterized by the $4$-velocity vector
\beq
n=\partial_t\,.
\eeq
An orthonormal spatial triad adapted to the observers $n \equiv e_0$ is given by\footnote{
Such a frame is also parallel propagated along $e_0$, i.e., $\nabla_{e_0}e_\alpha=0$.
}
\beq
e_1=\partial_z\,,\qquad
e_2=\frac{1}{F}\partial_x\,,\qquad
e_3=\frac{1}{G}\partial_y\,,
\eeq
with dual frame $n^\flat \equiv -\omega^0=-\rmd t$ and $\omega^1=\rmd z$, $\omega^2= F\, \rmd x$ and $\omega^3=G\, \rmd y$.
We find it convenient to introduce the  notation
\beq
e_+=e_2\otimes e_2+e_3\otimes e_3\,, \quad
e_\times=e_2\otimes e_3-e_3\otimes e_2=e_2\wedge e_3\,.
\eeq
The associated congruence of the observer world lines is geodesic and vorticity-free, but has a nonzero expansion.

In the case of an electromagnetic wave the metric (\ref{radfield}) depends on a single function ($F=G$), which will be denoted by $H$ below.

\subsection{Single electromagnetic plane waves}
\label{em1}

The metric (\ref{radfield}) associated with an electromagnetic plane wave is an electrovacuum solution  of the Einstein-Maxwell equations with
electromagnetic potential  $A^\flat$  and  Faraday tensor  $F^\flat =\rmd A^\flat $ given by
\beq
A^\flat=h(u)\,\rmd x\,,\qquad F=h'(u)\, \rmd u \wedge \rmd x\,.
\eeq
[Here a prime denotes differentiation with respect to $u$.]
The associated energy-momentum tensor is  that of a radiation field with flux factor $\Phi$, 
\beq
\label{Tmunuem}
T=\Phi^2 \partial_v \otimes \partial_v \,,\qquad 
\Phi=\sqrt{2}\frac{h'}{H}\,,
\eeq
and
the (nonvacuum) 
Einstein's field equations  
imply the (single) condition
\beq
\label{eqHh}
H'' + \frac{h'{}^2}{H}=0\,,   
\eeq
for the two unknown functions $H$ and $h$.
In order to determine $H$ and $h$ uniquely, one has to impose an additional relation between them.
For example, if $H$ is assumed to be  known (or, equivalently, if one fixes the background gravitational field),  Eq. (\ref{eqHh}) 
can be integrated exactly and the solution reads
\beq
\label{hsolgen}
h(u)=\int_0^u \sqrt{-H(x)H''(x)}\,\rmd x\,.
\eeq  
Eq. (\ref{hsolgen}) thus identifies a class of exact solutions of the Einstein-Maxwell field equations representing a plane electromagnetic wave. 
On the other hand, if one treats $h$ as the known function  (or, equivalently, if one fixes the background electromagnetic field), Eq. (\ref{eqHh}) is a second order differential equation for $H$, which cannot be solved in general, but only for special choices of $h$.

The frame components of $F$ are
\beq
F=\frac{\Phi}{2} [\omega^0 \wedge \omega^2 -\omega^1 \wedge \omega^2]
=n^\flat \wedge E(n) + {}^{*_{(n)}} B(n)\,,
\eeq
where the symbol $^{*_{(u)}}$ denotes the spatial dual of a spatial tensor with respect to $u$, i.e. taken with $\eta(u)_{\alpha\beta\gamma}=u^\mu \eta_{\mu\alpha\beta\gamma}$.
The electric and magnetic fields as measured by the fiducial observers $n$ are perpendicular to each other, have the same magnitude and nonzero components only on the wave front
\beq
E(n)=-\frac{\Phi}{2}\, e_2 \,,\qquad 
B(n)=\frac{\Phi}{2}\, e_3\,.
\eeq
The two electromagnetic invariants both vanish, as expected for a wave-like behavior of the electromagnetic field, i.e.,
\beq
E(n)^2-B(n)^2=0\,,\qquad 
E(n)\cdot B(n)=0\,.
\eeq
The nonzero frame components of the Riemann tensor are
\begin{eqnarray}
R_{0202}&=&R_{0303}=R_{1212}=R_{1313}\nonumber\\
&=&-R_{0313}=-R_{0212}\nonumber\\
&=&-\frac{H''}{2H}=\frac{\Phi^2}{4}\,,
\end{eqnarray}
so that its electric and magnetic parts are given by
\begin{eqnarray}
{\mathcal E}(n)&=&\frac{\Phi^2}{4}e_+
=E(n)\otimes E(n)+B(n)\otimes B(n)\,,\nonumber\\
{\mathcal H}(n)&=& -\frac{\Phi^2}{4} e_\times
=E(n)\wedge B(n)\,.
\end{eqnarray}
The electric and magnetic parts of the Weyl tensor are instead both vanishing, implying that the spacetime metric is conformally flat, and hence the associated gravitational field is algebraically special and of Petrov type O.
The Bel-Robinson tensor is then identically vanishing. 
The super-momentum tensor four vector ${\sf P}^{\rm (g)}_{\rm B}(u)$ built with the Bel tensor is a null vector and it is given by
\beq
{\sf P}^{\rm (g)}_{\rm B}(n)= -\left(\frac{H''}{H}\right)^2(e_0+e_1)
=-\frac{\Phi^4}{4}(e_0+e_1)\,,
\eeq
whereas that built with the Chevreton tensor is
\beq
{\sf P}^{\rm (g)}_{\rm C}(n)=-\frac{\left(\Phi' \right)^2}{4}(e_0+e_1)\,.
\eeq
${\sf P}^{\rm (g)}_{\rm C}(n)$ like ${\sf P}^{\rm (g)}_{\rm B}(n)$ is also a null vector.

Two different choices of $h$ which are of particular interest will be discussed below, corresponding to electromagnetic waves with either constant or oscillating profiles.

\subsubsection{Electromagnetic waves with constant profile}

The choice  $h(u)=\sin (bu)$, with $b$ constant
corresponds to the case of electromagnetic waves with constant profile \cite{Bell:1974vb} and implies a constant flux $\Phi=\sqrt{2}b$ of the associated radiation field and constant electric and magnetic fields as measured by the observers $n$, i.e.,
\beq
E(n)=-\frac{b}{\sqrt{2}} e_2 \,,\qquad 
B(n)=\frac{b}{\sqrt{2}} e_3\,,
\eeq
and $H(u)=\cos (bu)$.
Here $b$ (with  dimensions of the inverse of a length)  characterizes  the strength of the electromagnetic wave and it is related to the frequency of the wave by $b=\sqrt{2}\omega$.

The frame components of the electric and magnetic parts of the Riemann tensor are constant as well, namely
\beq
{\mathcal E}(n)= \frac{b^2}{2}e_+\,,\qquad
{\mathcal H}(n)= -\frac{b^2}{2} e_\times\,.
\eeq

The super-momentum tensor four vector is given by
\beq
{\sf P}^{\rm (g)}(n)= -b^4(e_0+e_1)\,,
\eeq
whereas the Chevreton tensor is identically vanishing.

\subsubsection{Electromagnetic waves with oscillating electric and magnetic fields}

Let us  choose the  function $h$ such that the electric and magnetic fields are both characterized by an oscillatory behavior, i.e., 
\beq
E(n)=-A \sin (bu) e_2 \,,\qquad 
B(n)=A \sin (bu) e_3\,,
\eeq
with $A$ and $b$ constants, by requiring
\beq
\label{fluxmath}
\frac{\Phi}{2}=\frac{h'}{\sqrt{2}H}=A\sin (bu)\,.
\eeq
Substituting then into Eq. (\ref{eqHh}) gives
\beq
\label{eqHhnew}
H''+2A^2\sin^2 (bu)H=0\,,
\eeq
which represents a Mathieu's differential equation.
A short review of main definitions and basic features of Mathieu functions is given in Appendix B of Ref. \cite{Bini:2014jba}, to which we also refer for notation and conventions.

Solving Eq. (\ref{eqHhnew}) with initial conditions $H(0)=1$ and $H'(0)=0$ yields
\begin{eqnarray}
\label{sol}
H(u) &=& {\rm MathieuC}\left(\frac{A^2}{b^2}, \frac{A^2}{2b^2}, bu\right)\,,\nonumber\\
h(u) &=& \int_0^u \sqrt{2}A\sin(b x)H(x)\,\rmd x\,.
\end{eqnarray}
The electric and magnetic parts of the Riemann tensor are also oscillating
\beq
{\mathcal E}(n)= A^2\sin^2(bu) e_+\,,\quad
{\mathcal H}(n)= -A^2\sin^2(bu) e_\times\,.
\eeq
The Bel and Chevreton super-momentum four vectors turn out to be
\beq
{\sf P}^{\rm (g)}(n)= -4A^4\sin^4(bu)(e_0+e_1)\,,
\eeq
and
\beq
{\sf P}^{\rm (g)}_{\rm C}(n)=-A^2b^2\cos^2(bu)(e_0+e_1)\,,
\eeq
respectively.

\subsection{Single gravitational plane waves}

Let us now consider the case of an exact gravitational plane wave.
Vacuum Einstein's field equations imply 
\beq
\label{eingw}
G''F+GF''=0\,.
\eeq
The Bel and Bel-Robinson tensors coincide in this case.
The super-momentum four vector is given by
\beq
{\sf P}^{\rm (g)}(n)= -\left(\frac{G''}{G}\right)^2(e_0+e_1)\,.
\eeq
A simple solution of Eq. (\ref{eingw}) is 
\beq
\label{FGgw}
F(u)=\cos(b_{\rm (gw)}u)\,, \qquad
G(u)=\cosh(b_{\rm (gw)}u)\,,
\eeq
where $\omega_{\rm (gw)}=b_{\rm (gw)}/\sqrt{2}$ is the frequency of the gravitational wave, so that the super-momentum tensor four vector reads
\beq
{\sf P}^{\rm (g)}(n)= -b_{\rm (gw)}^4(e_0+e_1)\,,
\eeq
which is a null vector as in the case of an electromagnetic wave.

\section{Colliding waves}

\begin{figure}
\begin{center}
\includegraphics[scale=0.55]{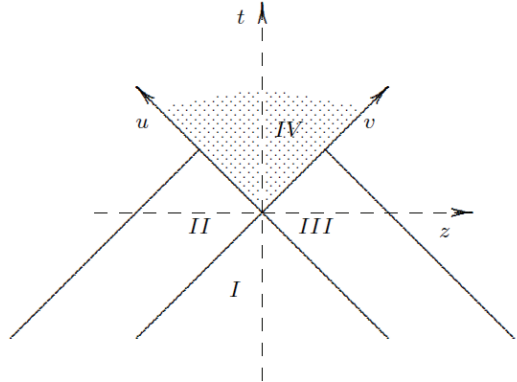}
\end{center}
\caption{\label{fig:1} The spacetime picture of two colliding plane waves. Region IV ($u\ge 0$ and $v\ge 0$) is the interaction region, Region II ($u\ge 0$ and $v\le 0$) and III ($u\le 0$ and $v\ge 0$) are single wave regions, Region I ($u\le 0$ and $v\le 0$) is the flat space corresponding to the situation before the passage of the waves.}
\end{figure}

Exact solutions of the Einstein equations representing colliding electromagnetic and gravitational plane waves have been discussed extensively in the literature \cite{Griffiths:1991zp}.
We will consider the simplest case of collinear polarization of the plane waves, so that the line element associated with the collision region has the general form \cite{Bell:1974vb,Griffiths:1975zm}
\beq
\label{metcollgen}
\rmd s^2=-2e^{M}\rmd u\rmd v+e^{-U}\left[e^V\rmd x^2+e^{-V}\rmd y^2\right]
\,,
\eeq
where all metric functions depend on $u$ and $v$, or equivalently $t$ and $z$, using Eq. (\ref{u_coord}).
An orthonormal frame adapted to a family of observers at rest with respect to the coordinates $(x,y,z)$ is given by
\beq\fl\qquad
\label{framecoll}
e_0=n=e^{-M/2}\partial_t\,,\quad
e_1=e^{-M/2}\partial_z\,,\quad
e_2=\frac{1}{\sqrt{g_{xx}}}\partial_x\,,\quad
e_3=\frac{1}{\sqrt{g_{yy}}}\partial_y\,.
\eeq
The spacetime associated with the whole collision process is graphically represented in Fig. \ref{fig:1}. It is a matching of different spacetime regions:  a flat spacetime region (of Petrov type-O, before the passage of the waves), two single wave spacetimes (Petrov type-N, corresponding to the two oppositely directed incoming waves) and a collision region (generally of Petrov type-I).
All the geometry can be illustrated in the $u$-$v$ plane (orthogonal to the plane wave fronts aligned with the $x$-$y$ planes), where one has the region IV (collision region) when the coordinates $u$ and $v$ vary in the range $u>0$, $v>0$, the regions II (where $u>0$ but $v<0$) and III  (where $u<0$ but $v>0$) corresponding to the two single wave regions and the flat space region I (where $u\le 0$ and $v\le 0$), before the passage of the waves.
Following Khan-Penrose \cite{Khan:1971vh}, one can then extend the metric (\ref{metcollgen}) from the collision region IV to whole spacetime by formally replacing 
\begin{equation}
\label{heav}
u\, \to\ u_+=u\, \theta(u)\,, \qquad 
v\, \to\ v_+=v\, \theta(v)\,,
\end{equation}
where $\theta(x)$ is the Heaviside step function. 
In this way the extended metric is continuous in general, but has discontinuous first derivatives along the null boundaries $u = 0$ and $v = 0$, so that the Riemann tensor acquires distributional-singular parts.


\subsection{Colliding electromagnetic plane waves}

As a model of colliding electromagnetic plane waves we consider the Bell-Szekeres solution \cite{Bell:1974vb}
\begin{eqnarray}
\label{collem}
\rmd s_{\rm IV}^2 &=& -2\rmd u \rmd v +\cos^2 (u-v) \rmd x^2 +\cos^2 (u+v) \rmd y^2
\,,
\end{eqnarray}
representing the interaction of two step-profile electromagnetic waves whose polarization vectors 
are aligned (see Section 3.1.1).
Note that we have set to unity the strength parameter of both the incoming waves, for simplicity.
The nonvanishing coordinate components of the Riemann tensor are 
\begin{eqnarray}
&&R_{uxux}=R_{vxvx}=-R_{uxvx}=\cos^2(u-v)\,,\nonumber\\
&&R_{uyuy}=R_{vyvy}=R_{uyvy}=\cos^2(u+v)\,.
\end{eqnarray}
The orthonormal frame (\ref{framecoll}) adapted to the static observers $n=e_0$ reads
\beq\fl\qquad
e_0=\partial_t\,,\qquad
e_1=\partial_z\,,\qquad
e_2=\frac{1}{\cos(u-v)}\partial_x\,,\qquad
e_3=\frac{1}{\cos(u+v)}\partial_y\,.
\eeq
The electric part of the Riemann tensor with respect to this frame is given by
\beq
{\mathcal E}(n)=2e_3\otimes e_3\,,
\eeq
whereas the magnetic part is identically vanishing ${\mathcal H}(n)=0$, implying that the spatial part of the gravitational super-momentum four vector vanishes too (see Eq. (\ref{Belsuperendef})).
The latter then turns out to be
\beq
{\sf P}^{\rm (g)}_{\rm IV}(n)=-4 \, e_0\,,
\eeq
i.e., timelike, and ${\mathcal E}^{\rm (g)}_{\rm IV}(n)=4$.

In region II, the single $u$-wave is described by the line element [extension of  the metric (\ref{collem}) by using the relations (\ref{heav})]
\beq
\rmd s_{\rm II}^2 = -2\rmd u \rmd v +\cos^2 u_+ (\rmd x^2 +\rmd y^2)\,,
\eeq
where $u_+=u\theta(u)$.  
The nonvanishing coordinate components of the Riemann tensor are 
\beq
R_{uxux}=R_{uyuy}=\cos^2u_+\theta(u)\,.
\eeq
The gravitational super-momentum four vector is then given by
\beq
{\sf P}^{\rm (g)}_{\rm II}(n)= -\theta(u)(e_0+e_1)  \,,
\eeq
i.e., ${\sf P}^{\rm (g)}_{\rm II}(n)$ is a null vector and ${\mathcal E}^{\rm (g)}_{\rm II}(n)=1$.

Similarly, in the $v$-region we get
\beq
{\sf P}^{\rm (g)}_{\rm III}(n)= -\theta(v)(e_0-e_1)  \,,
\eeq
with ${\mathcal E}^{\rm (g)}_{\rm III}(n)=1$, so that 
\beq
{\mathcal E}^{\rm (g)}_{\rm IV}(n)=[{\mathcal E}^{\rm (g)}_{\rm II}(n)+{\mathcal E}^{\rm (g)}_{\rm III}(n)]^2\,.
\eeq

One can then form a \lq\lq distributional" representation of the super-momentum vector in the whole spacetime as follows
\begin{eqnarray}
\label{Pgcollem}
{\sf P}^{\rm (g)}(n)&=& -[2\theta(u)\theta(v)+\theta(u)+\theta(v)]e_0+[-\theta(u)+\theta(v)]e_1 \nonumber\\
&=& -[\theta(u)+\theta(v)]^2e_0+[-\theta(u)+\theta(v)]e_1\,.
\end{eqnarray}

It is known that impulsive gravitational waves are created following the collision \cite{Griffiths:1991zp}, implying a redistribution of the energy of the incoming electromagnetic shock waves. In fact, by replacing $u\to u_+$ and $v\to v_+$ in Eq. (\ref{collem}) the Riemann tensor acquires terms proportional to the Dirac-delta function
\begin{eqnarray}
&&R_{uxux}=-\delta(2u)\sin 2v_++\theta(u)\cos^2(u_+-v_+)\,,\nonumber\\
&&R_{vxvx}=-\delta(2v)\sin 2u_++\theta(v)\cos^2(u_+-v_+)\,,\nonumber\\
&&R_{uxvx}=-\theta(u)\theta(v)\cos^2(u_+-v_+)\,,\nonumber\\
&&R_{uyuy}=\delta(2u)\sin 2v_++\theta(u)\cos^2(u_++v_+)\,,\nonumber\\
&&R_{vyvy}=\delta(2v)\sin 2u_++\theta(v)\cos^2(u_++v_+)\,,\nonumber\\
&&R_{uyvy}=\theta(u)\theta(v)\cos^2(u_++v_+)\,,
\end{eqnarray}
i.e., formally
\begin{eqnarray}
\label{eq-formal}
R_{\alpha\beta\gamma\delta} &=&R_{\alpha\beta\gamma\delta}^{{\rm sing} } +R_{\alpha\beta\gamma\delta}^{ \rm II}\theta(u)+R_{\alpha\beta\gamma\delta}^{ \rm III}\theta(v)+R_{\alpha\beta\gamma\delta}^{\rm IV}\theta(u)\theta(v) \,,
\end{eqnarray}
with
\beq
R_{\alpha\beta\gamma\delta}^{{\rm sing} } = R_{\alpha\beta\gamma\delta}^{u-{\rm sing}}\delta(u)+R_{\alpha\beta\gamma\delta}^{v-{\rm sing}}\delta(v)\,.
\eeq
The total super-momentum four vector (\ref{Pgcollem}) thus becomes
\beq
{\sf P}^{\rm (g)}(n)={\sf P}^{\rm (g)}_{\rm reg}(n)+{\sf P}^{\rm (g)}_{\rm sing}(n)\,,
\eeq
where the regular part and the singular part are given by 
\beq
{\sf P}^{\rm (g)}_{\rm reg}(n)=-[\theta(u)+\theta(v)]^2e_0+[-\theta(u)+\theta(v)]e_1\,,
\eeq
and 
\beq
{\sf P}^{\rm (g)}_{\rm sing}(n)=-[\delta(u)\tan v_++\delta(v)\tan u_+]e_0\,,
\eeq
respectively.

The Faraday tensor is given by
\beq
F=\frac{1}{\sqrt{2}} \{[\theta(u)-\theta(v)]\omega^0 \wedge \omega^2 -[\theta(u)+\theta(v)]\omega^1 \wedge \omega^2\}\,.
\eeq
Taking the covariant derivative generates $\delta$ function terms, so that the Chevreton tensor will contain terms proportional to the square of $\delta$, thus becoming even more singular than the Bel tensor at the boundaries of the interaction region, whereas it is identically vanishing in the single wave regions.
The Chevreton super-momentum four vector then turns out to be completely singular, involving squares of Dirac-delta functions
\beq
{\sf P}^{\rm (g)}_{\rm C}(n)=-\frac12[\delta(u)+\delta(v)]e_0-\frac12[\delta^2(u)-\delta^2(v)]e_1\,.
\eeq

\subsection{Colliding gravitational plane waves}
\label{gw1}

The spacetime geometry associated with two colliding gravitational plane waves is characterized by the presence of either a spacetime singularity or a Killing-Cauchy horizon as a result of the nonlinear wave interaction \cite{Khan:1971vh,Ferrari:1987yk,Ferrari:1987cs}. 
We consider below a type-D solution belonging to the Ferrari-Ibanez class of solutions \cite{Ferrari:1988nu}, with line element 
\begin{eqnarray} 
\label{summa}
\rmd s^2 
 &=& -4S_-^2 \rmd u\rmd v  +\frac{S_+}{S_-}\rmd x^2+C^2 S_-^2\rmd y^2\,,
\end{eqnarray}
where we have used the notation
\beq
S_\pm=S_\pm(u+v) = 1\pm \sin(u+v)\,, \qquad C=\cos(u-v)\,.
\eeq
This metric develops a singularity at $u+v=\pi/2$, where the Kretschmann curvature invariant
\beq
R_{\alpha\beta\gamma\delta}R^{\alpha\beta\gamma\delta}
=\frac{48}{S_-^6}
\eeq
diverges.
The nonvanishing components of the Riemann tensor in the interaction region are the following
\beq
\begin{array}{lll}
R_{vxvx}=R_{uxux}=3R_{uxvx}= -3\displaystyle\frac{S_+}{S_-^2}\,,\quad  &  R_{xyxy}= -C^2\displaystyle\frac{S_+}{S_-^2}\,,\quad   &  
\cr
&& \cr
R_{vyvy}=R_{uyuy}=-3R_{uyvy}= 3C^2S_-  \,,\quad   &  R_{uvuv}=4S_-    \,.\quad     &   
\cr
&& \cr
\end{array}
\eeq
The orthonormal frame (\ref{framecoll}) adapted to the static observers $n=e_0$ reads
\beq\fl\qquad
e_0=\frac{1}{\sqrt{2}S_-}\partial_t\,,\quad
e_1=\frac{1}{\sqrt{2}S_-}\partial_z\,,\quad
e_2=\left(\frac{S_+}{S_-}\right)^{-1/2}\partial_x\,,\quad
e_3=\frac{1}{CS_-}\partial_y\,.
\eeq
The electric part of the Riemann tensor with respect to this frame is given by
\beq
{\mathcal E}(n)=\frac{1}{S_-^3}[e_1\otimes e_1-2e_2\otimes e_2+e_3\otimes e_3]\,,
\eeq
whereas the magnetic part is identically vanishing ${\mathcal H}(n)=0$, implying that the spatial part of the Bel super-momentum four vector vanishes too (see Eq. (\ref{Belsuperendef})).
The latter then turns out to be
\beq
\label{PBgw4}
{\sf P}^{\rm (g)}_{\rm IV}(n)=-\frac{6}{S_-^6}  e_0
=-{\mathcal E}^{\rm (g)}_{\rm IV}(n)\,  e_0\,,
\eeq
and is timelike.

As stated above, replacing in region IV $u\to u_+$ and $v\to v_+$ allows the extension of the metric to the whole spacetime 
\begin{eqnarray}
\label{metall}
\rmd s_{\rm I}^2&=&-4\rmd u\rmd v+\rmd x^2+\rmd y^2\,, \nonumber\\
\rmd s_{\rm II}^2&=&-4S_-^2(u_+) \rmd u\rmd v+\frac{S_+(u_+)}{S_-(u_+)}\rmd x^2+\cos^2u_+ S_-^2(u_+)\rmd y^2\,, \nonumber\\
\rmd s_{\rm III}^2&=&-4S_-^2(v_+) \rmd u\rmd v+\frac{S_+(v_+)}{S_-(v_+)}\rmd x^2+\cos^2v_+ S_-^2(v_+)\rmd y^2\,.
\end{eqnarray}
The structure of the Riemann tensor components in the collision region is analogous to the one given by Eq. (\ref{eq-formal}), with rather more involved expressions which is not necessary to show explicitly.

In the single $u$-wave region the nonvanishing components of the Riemann tensor are
\begin{eqnarray}
R_{uxux}&=& -\delta(u)-3\frac{S_+(u_+)}{S_-^2(u_+)}\theta(u)\,,\nonumber\\
R_{uyuy}&=& \delta(u)+3\cos(u_+)^2S_-(u_+)\theta(u)\,,
\end{eqnarray}
implying that the initial approaching wave is a combination of impulsive and shock waves.
The Bel super-momentum four vector is thus singular at the boundary
\begin{eqnarray}
\label{PBgw2}
{\sf P}^{\rm (g)}_{\rm II}(n)&=&-\left[\frac14\delta^2(u)+\frac32\delta(u)+\frac{9}{4}\frac{\theta(u)}{S_-^6(u_+)}\right]  (e_0+e_1)\nonumber\\
&=&-{\mathcal E}^{\rm (g)}_{\rm II}(n)\, (e_0+e_1)\,,
\end{eqnarray}
and is a null vector.
Similar considerations hold for the single $v$-wave region, where 
\begin{eqnarray}
\label{PBgw3}
{\sf P}^{\rm (g)}_{\rm III}(n)&=&-\left[\frac14\delta^2(v)+\frac32\delta(v)+\frac{9}{4}\frac{\theta(v)}{S_-^6(v_+)}\right]  (e_0-e_1)\nonumber\\
&=&-{\mathcal E}^{\rm (g)}_{\rm III}(n)\, (e_0-e_1)\,.
\end{eqnarray}

Performing the extension to the whole spacetime, the total super-momentum four vector reads
\beq
{\sf P}^{\rm (g)}(n)={\sf P}^{\rm (g)}_{\rm reg}(n)+{\sf P}^{\rm (g)}_{\rm sing}(n)\,,
\eeq
where the regular part is given by 
\begin{eqnarray}
{\sf P}^{\rm (g)}_{\rm reg}(n)&=&-\frac{9}{4S_-^6(u_++v_+)}\left\{\left[\theta(u)+\theta(v)+\frac23\theta(u)\theta(v)\right]e_0\right.\nonumber\\
&&\left.
+[\theta(u)-\theta(v)]e_1\right\}\,,
\end{eqnarray}
and the singular part ${\sf P}^{\rm (g)}_{\rm sing}(n)$ contains terms proportional to the Dirac functions $\delta(u)$, $\delta(v)$, their product and their square.

The relation between the super-energy in the collision region and in the single wave regions is more involved than in the electromagnetic case, as expected.
This can be seen, for example, by eliminating $u_+$ and $v_+$ from the regular parts of Eqs. (\ref{PBgw2})--(\ref{PBgw3}) and substituting in Eq. (\ref{PBgw4}), so that
\beq
{\mathcal E}^{\rm (g)}_{\rm IV}(n)= 
\frac{6}{\left[1- \sin\left(\arcsin \alpha_{\rm II}(n)+\arcsin \alpha_{\rm III}(n)\right)\right]^6}\,,
\eeq
with
\begin{eqnarray}
\alpha_{\rm II, III}(n) &=&  -\left(\frac{9}{4{\mathcal E}^{\rm (g)}_{\rm II, III}(n)} \right)^{1/6}-1 \,.
\end{eqnarray}

\subsection{The role of the observer in the collision region}

Up to now we have only mentioned the dependence of the super energy-momentum four vectors on the choice of the observer and used instead natural observers, i.e., at rest  with respect to a set of Cartesian-like coordinate system which can be associated with all the various spacetime regions involved. 
This choice is indeed a natural choice, which facilitates interpretation and help the intuition: an observer at rest with respect to the (spatial) coordinates in the $u$-wave region  sees a wave propagating along the positive $z$ direction, while in the $v$-wave sees a wave propagating in the opposite direction; in the collision region this observer is actually the one lying in \lq\lq the center of momenta" of the system of two waves, in the sense that he sees \lq\lq symmetrically" the two waves approaching each other\footnote{
This is true only if the two incoming waves have been taken as symmetric, in the sense that they have the same geometrical profile, as it has been the case in the present study.
}.  

This privileged observer can be actually embedded in a family of observers which are in motion along the $z$-axis with four velocity 
\beq
\label{Udef}
U=\gamma (n +\nu e_1)\,, \qquad
\gamma=(1-\nu^2)^{-1/2}\,.
\eeq
Taking $E_0=U$, an adapted spatial triad to $U$ is given by
\beq
E_1=\gamma (\nu n +e_1)\,,\qquad E_2=e_2\,,\qquad E_3=e_3\,.
\eeq
The electric and magnetic parts of the Riemann tensor with respect to this frame are given by
\begin{eqnarray}\fl\qquad
{\mathcal E}(U)&=&\frac{\gamma^2}{S_-^3}[(1-\nu^2)E_1\otimes E_1-2(1+\nu^2)E_2\otimes E_2+(1+2\nu^2)E_3\otimes E_3]
\,, \nonumber\\
\fl\qquad
{\mathcal H}(U)&=&-\frac{3\gamma^2\nu}{S_-^3}[E_2\otimes E_3+E_3\otimes E_2]\,,
\end{eqnarray}
so that the spatial part of the Bel super-momentum four vector is different from zero in general, unless $\nu=0$. 
In fact, the latter turns out to be
\beq
{\sf P}^{\rm (g)}_{\rm IV}(U)=-\frac{6}{S_-^6}\gamma^4 [(1+4\nu^2+\nu^4)E_0-3\nu(1+\nu^2)E_1]\,,
\eeq
so that the $\nu$-dependent gravitational super-energy content is given by
\beq
{\mathcal E}^{\rm (g)}_{\rm IV}(U)= \frac{6}{S_-^6}\gamma^4(1+4\nu^2+\nu^4)\,,
\eeq 
and it is minimum when $\nu=0$. In fact, when $\nu=0$ the observer $U$ reduces to $n$ and sees symmetrically the collision process, the spatial part of the super-momentum being zero due to the vanishing of the magnetic part of the Riemann tensor.
In this respect, the family of static observers in the collision region play a role similar to the Carter's family of observers in black hole spacetimes, who measure aligned electric and magnetic parts of the Riemann tensor, so that their cross product is identically vanishing (see Eq. (\ref{Belsuperendef})) \cite{bini-jan-miniutti}. In this sense we should have denoted $e_0=n=u_{\rm (car)}$, and characterized $e_0$ so far from the physical point of view of being associated with  \lq\lq minimal super-energy observers," instead of their geometric, coordinate adapted characterization of observers at rest with respect to the spatial coordinates.  

Let us turn to the general case $\nu\not=0$. 
The spatial super-momentum is nonzero and given by
\beq
{\mathcal P}^{\rm (g)}_{\rm IV}(U)= \frac{18}{S_-^6}\gamma^4\nu(1+\nu^2)E_1
={\mathcal P}^{\rm (g)}_{\rm IV}(U)^1\,E_1\,,
\eeq
i.e., according to the observer $U$, a nonzero contribution to the super-momen\-tum is carried by the waves in the $E_1$ (boosted $z$) direction, which disappears when $\nu=0$ and spans all real values (from $-\infty$ to $+\infty$) as soon as $\nu$ varies from $-1<\nu<1$.

\begin{figure}
\begin{center}
\includegraphics[scale=.35]{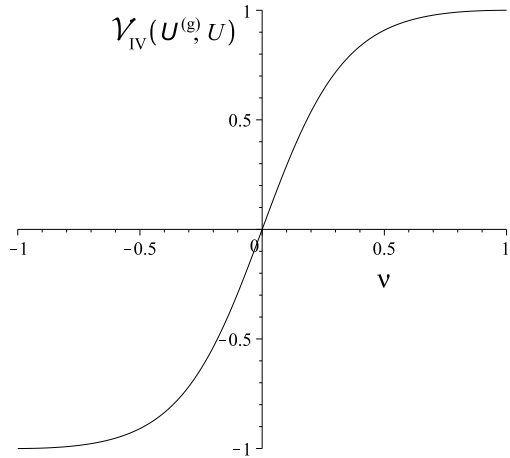} 
\end{center}
\caption{The behavior of the velocity field ${\mathcal V}_{\rm IV}({\sf U}^{\rm (g)},U)$ in the interaction region (see Eq. (\ref{vcoll})) is shown as a function of the linear velocity $\nu\in(-1,1)$ for an observer moving along the $z$-direction with constant speed. 
It remains constant for every fixed value of $\nu$ during the whole collision process.
}
\label{fig:2}
\end{figure}

\begin{figure}
\begin{center}
\includegraphics[scale=0.35]{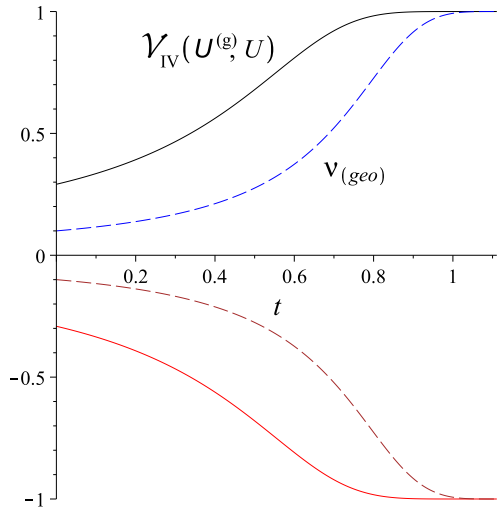}
\end{center}
\caption{The behavior of the velocity field ${\mathcal V}_{\rm IV}({\sf U}^{\rm (g)},U)$ in the interaction region (see Eq. (\ref{vcoll})) for a geodesic observer moving along the $z$-direction is shown as a function of $t=(u+v)/\sqrt{2}$ for $\nu_0=0.1$ (solid curves). Dashed curves show instead the behavior of the observer's linear velocity $\nu_{\rm (geo)}$ (see Eq. (\ref{nugeocoll})).
Both of them reach the values $\pm1$ at the end of the collision process ($t=\pi/2\sqrt{2}\approx1.11$).
}
\label{fig:3}
\end{figure}

In order to qualitatively describe the features of the nonlinear interaction of the radiative fields associated with the waves it is interesting to study the behavior of the velocity field (\ref{calVdef}), i.e., the relative velocity of the unit timelike vector ${\sf U}^{\rm (g)}$ aligned with the super-momentum with respect to the reference observer $U$ \cite{Ibanez:1985ps}. Note that this quantity is positive definite, being the ratio of the magnitude of the spatial super-momentum vector and the super-energy density.  However, since the spatial super-momentum has only one nonvanishing frame component along the direction of propagation of the incoming waves, one can allow the relative velocity to take both signs. 
In the collision region we get 
\beq
\label{vcoll}
{\mathcal V}_{\rm IV}({\sf U}^{\rm (g)},U)=\frac{{\mathcal P}^{\rm (g)}_{\rm IV}(U)^1}{{\mathcal E}^{\rm (g)}_{\rm IV}(U)}
=\frac{3\nu(1+\nu^2)}{1+4\nu^2+\nu^4}\,.
\eeq
Extending this definition to the single-wave regions, we have that before the collision the velocity field is ${\mathcal V}_{\rm II}({\sf U}^{\rm (g)},n)=1$ and ${\mathcal V}_{\rm III}({\sf U}^{\rm (g)},n)=-1$ (see Eqs. (\ref{PBgw2})--(\ref{PBgw3})), according to the fact that in those regions the gravitational field is purely radiative and the waves are moving at the speed of light.
Furthermore, there is no need to have an observer moving along $z$ in a single wave regions, because there is not any \lq\lq symmetric" situation to look at.
In the interaction region, instead, ${\mathcal V}_{\rm IV}({\sf U}^{\rm (g)},U)$ can attain any value between $-1$ and $1$. For instance, in the case of an observer moving with constant speed $\nu$, the velocity field (\ref{vcoll}) is also constant for fixed $\nu$ during the whole collision process.
Its behavior as a function of $\nu$ is shown in Fig. \ref{fig:2}.
A more interesting (and realistic) case is that of a geodesic observer, whose four velocity is given by Eq. (\ref{Udef}) with 
\beq
\label{nugeocoll}
\nu_{\rm (geo)}=\pm\left(1+\frac{S_-^2(u+v)}{\gamma_0^2\nu_0^2}\right)^{-1}\,,
\eeq 
where $\nu_0=\nu_{\rm (geo)}(0)$, and $\nu_{\rm (geo)}\to\pm1$ as the singularity $u+v=\pi/2$ is approached.
Its behavior as a function of $t=(u+v)/\sqrt{2}$ is shown in Fig. \ref{fig:3} for $\nu_0=0.1$, as an example.
Fig. \ref{fig:3} also shows the corresponding behavior of the relative velocity (\ref{vcoll}), which goes to $\pm1$ as well at the end of the collision process.
Note that this is not a general feature of colliding gravitational wave spacetimes. In fact, for background metrics not developing a singularity as a result of the nonlinear interaction of the waves the square of this relative velocity remains less that one, so that the gravitational field is no longer purely radiative \cite{Breton:1993du}. 

A similar discussion can be clearly done also in the case of collision of electromagnetic waves.
The observer (\ref{Udef}) moving  along the $z$-axis with constant speed is also geodesic in this case.
The electric and magnetic parts of the Riemann tensor are given by
\begin{eqnarray}
{\mathcal E}(U)&=&2\gamma^2[\nu^2E_2\otimes E_2+E_3\otimes E_3]
\,, \nonumber\\
{\mathcal H}(U)&=&-2\gamma^2\nu[E_2\otimes E_3-E_3\otimes E_2]\,,
\end{eqnarray}
so that the Bel super-momentum four vector turns out to be
\beq
{\sf P}^{\rm (g)}_{\rm IV}(U)=-4\gamma^4(1+\nu^2)^2\left[E_0-\frac{2\nu}{1+\nu^2}E_1\right]\,.
\eeq
The relative velocity (\ref{calVdef}) is given by
\beq
{\mathcal V}_{\rm IV}({\sf U}^{\rm (g)},U)=\frac{2\nu}{1+\nu^2}\,.
\eeq
Its behavior as a function of $\nu$ is quite similar to that shown in Fig. \ref{fig:2}.


\section{Conclusions}

We have studied the energy content of (exact) electromagnetic and gravitational plane wave spacetimes in terms of super-energy tensors, namely the Bel and Bel-Robinson tensors (defined in both cases) and the  Chevreton tensor (defined only in the electromagnetic case) with respect to natural observers.
The super-energy density and the spatial super-momentum density defined through these tensors have been combined into a single super-momentum four vector, which is an observer-dependent covariant quantity.

We have considered different situations corresponding to both single and colliding electromagnetic and gravitational waves.
In the case of single waves traveling along a preferred axis a natural observer family is that of observers at rest with respect to the spatial coordinates. 
The spacetime associated with electromagnetic plane waves with a single polarization state is conformally flat, implying that the Weyl tensor and then the Bel-Robinson tensor are identically vanishing.
The gravitational super-energy built with the Bel tensor turns out to be proportional to the fourth power of the radiation flux, whereas the electromagnetic one built with the Chevreton tensor is proportional to its derivative squared, implying that the Chevreton super-energy density vanishes for constant flux. 
Gravitational plane wave spacetimes have, instead, vanishing Chevreton tensor, while the Bel and Bel-Robinson tensors coincide.
The associated super-momentum four vectors are null in both cases.

Passing from a single wave region to the collision one the super-momentum four vector changes its causality property, becoming a timelike vector aligned with the observer's four velocity due to the nonlinear interaction of the incoming waves. [A possible, qualitative interpretation of this situation is that --because of the high-energy scattering of the two waves-- in the process enough energy is available to create a massive particle, or a black hole, as a product of the collision.]
Furthermore, as a result of the extension of the metric from the collision region to the whole spacetime matching the single-wave and flat regions, the super-momentum four vectors acquire singular parts at the boundaries, where impulsive gravitational waves are created. 
We have found that the regular part of the Bel super-energy in the collision region is simply the square of the sum of the super-energies of the two single waves in the electromagnetic case, whereas for gravitational waves it is nontrivially related to the super-energies of the two single waves.
The Chevreton super-momentum four vector is instead highly singular.

Finally, we have investigated the role of the observer by considering a family of observers moving along the direction of propagation of the incoming waves in the collision region. We have shown that the static observers are special within this family because they measure a minimal super-energy density.
The spatial part of the gravitational super-momentum four vector has only a nonvanishing component along the direction of propagation of the waves, whose magnitude diverges as the singularity is approached at the end of the collision process. The super-energy density is also diverging in this limit.
Therefore, it proves useful to study the behavior of their ratio, being the relative velocity of the unit timelike vector aligned with the super-momentum with respect to the reference observer. For instance, in the case of geodesic observers this velocity goes to unity for colliding gravitational waves, being a signature of the existence of a spacetime singularity rather than a Killing-Cauchy horizon.

\appendix

\section{Relation between the Bel and Bel-Robinson super-energy and super-momentum densities}

The orthogonal decomposition of the Bel and Bel-Robinson tensors with respect to the observer congruence with unit tangent vector $u$ has been given in Ref. \cite{GomezLobo:2007gm}. We will write here explicitly the observer-dependent expressions of the Bel and Bel-Robinson super-energy and super-momentum densities as well as the relation between them.

Let $\{e_\alpha\}$ be an adapted frame to $u$ (not necessarily orthonormal), so that $u=e_0$ and $\{e_a\}$ ($a=1,2,3$) is a spatial triad in the local rest space of $u$ \cite{Jantzen:1992rg}. 
Tensors and tensor fields with no components along $u$ are called spatial
with respect to $u$.

The Riemann tensor is decomposed in terms of the following three independent spatial tensor fields
\begin{eqnarray}
{\mathcal E}(u)_{ab}&=&R_{0a0b}\,, \nonumber \\
{\mathcal H}(u)_{ab}&=&R^*{}_{0a0b}=\frac12 R_{0acd}\eta(u)^{cd}{}_b\,, \nonumber \\
{\mathcal F}(u)_{ab}&=& {}^*R^*{}_{0a0b}=\frac14 R_{cdef}\eta(u)^{cd}{}_a \eta(u)^{ef}{}_b \,,
\end{eqnarray}
which are referred to as its electric, magnetic and mixed parts, respectively, with $\eta(u)_{abc}=\eta_{0abc}$.
In four-dimensional matrix notation we can equivalently write
\beq
\fl\quad
R_{\alpha \mu \beta \nu}u^\alpha u^\beta=
\left(
\begin{array}{c|c}
0 & 0 \cr
\hline
0 & {\mathcal E}(u) 
\end{array}
\right)
\,,\qquad 
{}^*R_{\alpha \mu \beta \nu}u^\alpha u^\beta=
\left(
\begin{array}{c|c}
0 & 0 \cr
\hline
0 & {\mathcal H}^{\rm T}(u) 
\end{array}
\right)
\,,
\eeq
and
\beq
\fl\quad
R^*{}_{\alpha \mu \beta \nu}u^\alpha u^\beta=
\left(
\begin{array}{c|c}
0 & 0 \cr
\hline
0 & {\mathcal H}(u) 
\end{array}
\right)
\,,\qquad 
{}^*R^*{}_{\alpha \mu \beta \nu}u^\alpha u^\beta=
\left(
\begin{array}{c|c}
0 & 0 \cr
\hline
0 & {\mathcal F}(u) 
\end{array}
\right)\,,
\eeq
assuming frame components.
In the following we will omit the explicit dependence on the observer $u$ to ease notation, i.e.,  ${\mathcal E}(u)={\mathcal E}$, etc..

Let us consider first the Bel tensor (\ref{Bel}) with associated super-energy and super-momentum densities (\ref{Belsuperendef}).
The full orthogonal splitting of the Bel tensor can be found in Ref. \cite{GomezLobo:2007gm} (see Eqs. (6.1)--(6.4) there).
Working with three-dimensional matrices and (spatial) frame indices, one finds for the super-energy \cite{bel58,bel59,bel62,rob59}
\begin{eqnarray}
\label{Belsuperendef2}
 k^{-1}\mathcal{E}^{\rm{(g)}}_{\rm B}(u)
 &=&  {\rm Tr}[{\mathcal E}]^2+{\rm Tr}[{\mathcal F}]^2 +2 {\rm Tr}[{\mathcal H}{\mathcal H}^{\rm T}]  \ ,
 \end{eqnarray}
(being $ {\rm Tr}[{\mathcal H}{\mathcal H}^{\rm T}]= {\rm Tr}[{\mathcal H}^{\rm T}{\mathcal H}]$), 
while the (frame) components of the super-momentum turn out to be
\beq
\label{Belsupermomdef2}
k^{-1}[\mathcal{P}^{\rm{(g)}}_{\rm B} (u)]^a =2 \eta^{abc}[{\mathcal E}{\mathcal H}-{\mathcal F}{\mathcal H}^{\rm T}]_{bc} \,,
\eeq
or, explicitly,
\begin{eqnarray}
\fl\qquad
k^{-1}[\mathcal{P}^{\rm{(g)}}_{\rm B} (u)]_1 &=& 2\left([{\mathcal E}{\mathcal H}]_{23}-
[{\mathcal E}{\mathcal H}]_{32}
\right)- 2\left([{\mathcal F}{\mathcal H}^{\rm T}]_{23}-
[{\mathcal F}{\mathcal H}^{\rm T}]_{32}
\right)\,,
\nonumber\\
\fl\qquad
k^{-1}[\mathcal{P}^{\rm{(g)}}_{\rm B} (u)]_2 &=&  -2\left([{\mathcal E}{\mathcal H}]_{13}-
[{\mathcal E}{\mathcal H}]_{31}
\right)+ 2\left([{\mathcal F}{\mathcal H}^{\rm T}]_{13}-
[{\mathcal F}{\mathcal H}^{\rm T}]_{31}
\right)\,,
\nonumber\\
\fl\qquad
k^{-1}[\mathcal{P}^{\rm{(g)}}_{\rm B} (u)]_3 &=&  2\left([{\mathcal E}{\mathcal H}]_{12}-
[{\mathcal E}{\mathcal H}]_{21}
\right) 
-
2\left([{\mathcal F}{\mathcal H}^{\rm T}]_{12}-
[{\mathcal F}{\mathcal H}^{\rm T}]_{21}
\right) \,.
\end{eqnarray}
In the vacuum case ${\mathcal F}=-{\mathcal E}$ and ${\mathcal H}^{\rm T}={\mathcal H}$, so that Eqs. (\ref{Belsuperendef2})--(\ref{Belsupermomdef2}) reduce to \cite{bel58,bel59,bel62,rob59}
\begin{eqnarray}
\label{Belsuperenmomvacuum}
k^{-1}\mathcal{E}^{\rm{(g)}}_{\rm B}(u)
&=&2 \left( {\rm Tr}[{\mathcal E}]^2  + {\rm Tr}[{\mathcal H}]^2\right)\,,\nonumber\\
k^{-1}[\mathcal{P}^{\rm{(g)}}_{\rm B} (u)]^a &=&4 \eta^{abc}[{\mathcal E}{\mathcal H}]_{bc} \,.
\end{eqnarray}

Consider then the Bel-Robinson tensor (\ref{BelR}). 
The Weyl tensor can be split into its electric and magnetic parts (\ref{EHweyl}), due to its self-duality property (i.e., ${}^*C^*{}=-C$), namely 
\beq
E(u)_{ab}=C_{0a0b}\,,\qquad 
H(u)_{ab}=C^*{}_{0a0b}\,.
\eeq
The latter are related to the electric, magnetic and mixed parts of the Riemann tensor by
\begin{eqnarray}
E  &=&\frac12 [{\mathcal E}-{\mathcal F}]^{\rm TF} \,,\qquad 
H  ={\rm SYM} [{\mathcal H}]\,,
\end{eqnarray}
where TF denotes the trace-free part of a tensor, and 
\beq
{\rm SYM} [{\mathcal H}]= \frac12 ({\mathcal H}+{\mathcal H}^{\rm T}), \qquad
{\rm ALT} [{\mathcal H}]= \frac12 ({\mathcal H}-{\mathcal H}^{\rm T}) \,,
\eeq
or in components
\beq\fl\qquad
E^a{}_b= \frac12 \left[{\mathcal E}^a{}_b-{\mathcal F}^a{}_b-\frac13\delta^a{}_b ({\rm Tr}[{\mathcal E}]-{\rm Tr}[{\mathcal F}])\right]\,,\qquad
H^{ab}= {\mathcal H}^{(ab)}\, .
\eeq
The full orthogonal splitting of the Bel-Robinson tensor can be found in Ref. \cite{GomezLobo:2007gm} (see Eqs. (5.1)--(5.4) there).
The associated super-energy and super-momentum densities (\ref{BelRsuperendef}) are given by
\begin{eqnarray}
\label{BelRsuperenmomdef2}
k^{-1}\mathcal{E}^{\rm{(g)}}_{\rm BR}(u)
&=&2 \left( {\rm Tr}[E]^2  + {\rm Tr}[H]^2\right)\,,\nonumber\\
k^{-1}[\mathcal{P}^{\rm{(g)}}_{\rm BR} (u)]^a &=&4 \eta^{abc}[EH]_{bc} \,,
\end{eqnarray}
respectively.
The latter reduce to Eq. (\ref{Belsuperenmomvacuum}) in vacuum, where the Weyl and Riemann tensors coincide.

One can use the relation between the Weyl and Riemann tensors to express the Bel super-energy and super-momentum densities (\ref{Belsuperendef2}) and (\ref{Belsupermomdef2}) in terms of the corresponding quantities (\ref{BelRsuperenmomdef2}) built up with the Bel-Robinson tensor.
In fact, the Riemann tensor can be written in terms of the Weyl tensor, the Ricci tensor and the scalar curvature as follows
\beq
R^{\alpha\beta}{}_{\gamma\delta}=C^{\alpha\beta}{}_{\gamma\delta}+2Q^{[\alpha}{}_{[\gamma }\delta^{\beta]}{}_{\delta ]}+\frac{1}{12}R \delta^{\alpha\beta}{}_{\gamma\delta}\,,
\eeq 
where
\beq
Q^\alpha{}_\beta =R^\alpha{}_\beta -\frac14 \delta^\alpha{}_{\beta} R\,,
\eeq
is the tracefree part of the Ricci tensor.
The spatial fields which represent the Ricci tensor are related to the electric and magnetic parts of the Riemann curvature tensor as
\beq
\fl\qquad
R^a{}_b = -{\mathcal E}^a{}_b-{\mathcal F}^a{}_b+\delta^a{}_b {\rm Tr}[{\mathcal F}]\,,\quad 
R^0{}_a =\eta_{abc}{\mathcal H}^{bc}\,,\quad
R^0{}_0 = -{\rm Tr}[{\mathcal E}]\,.
\eeq
The Bel and Bel-Robinson super-energy and super-momentum densities are related by (see also Refs. \cite{Bonilla:1997,Senovilla:1999xz})
\begin{eqnarray}
\mathcal{E}^{\rm{(g)}}_{\rm BR}(u) -\mathcal{E}^{\rm{(g)}}_{\rm B}(u)&=&\mathcal{E}^{\rm{(g)}}_{\rm matter}(u)
\,,\nonumber\\
{}[\mathcal{P}^{\rm{(g)}}_{\rm BR} (u)-\mathcal{P}^{\rm{(g)}}_{\rm B} (u)]^a&=&[\mathcal{P}^{\rm{(g)}}_{\rm matter} (u)]^a\,,
\end{eqnarray}
where
\begin{eqnarray}
k^{-1} \mathcal{E}^{\rm{(g)}}_{\rm matter}&=& 2{\rm Tr}[({\rm ALT} [{\mathcal H}])^2]\nonumber\\
&-&\frac12 \left({\rm Tr}[{\mathcal E}]^2+\frac13  ({\rm Tr}[{\mathcal E}])^2\right)\nonumber\\
&-& \frac12 \left( {\rm Tr}[{\mathcal F}]^2+\frac13 ({\rm Tr}[{\mathcal F}])^2\right)\nonumber\\
&-& {\rm Tr}[{\mathcal E}{\mathcal F}]+\frac13 {\rm Tr}[{\mathcal E}]{\rm Tr}[{\mathcal F}]\,,
\end{eqnarray}
and
\beq\fl\qquad
k^{-1} [\mathcal{P}^{\rm{(g)}}_{\rm matter} (u)]^a= 2\left[(\mathcal{E}+\mathcal{F})\,{\rm ALT} [{\mathcal H}]+\frac13({\rm Tr}[{\mathcal E}]-{\rm Tr}[{\mathcal F}])\,{\rm SYM} [{\mathcal H}]\right]^a\,.
\eeq
In vacuum  ${\rm ALT} [{\mathcal H}]=0$, ${\mathcal E}=-{\mathcal F}$ and ${\rm Tr}[{\mathcal E}]=0={\rm Tr}[{\mathcal F}]$, implying that $\mathcal{E}^{\rm{(g)}}_{\rm matter}=0=[\mathcal{P}^{\rm{(g)}}_{\rm matter} (u)]^a$, and hence $\mathcal{E}^{\rm{(g)}}_{\rm BR}(u)=\mathcal{E}^{\rm{(g)}}_{\rm B}(u)$ and $[\mathcal{P}^{\rm{(g)}}_{\rm BR} (u)]^a=[\mathcal{P}^{\rm{(g)}}_{\rm B} (u)]^a$.

\section*{Acknowledgements}

D.B. thanks Profs. R.T. Jantzen, B. Mashhoon and K. Ros\-quist for useful discussions about related topics at various stages during the development of the present project.  
We are indebted to Prof. J.M.M. Senovilla for a careful reading of the manuscript and valuable comments, and a clarification of the Bel and Bel-Robinson tensor formulas.

\end{document}